\documentstyle[12pt]{article}
\renewcommand{\theequation}{\thesection.\arabic{equation}}
\newcommand{\changeeq}{  
\setcounter{enumi}{\value{equation}}
\addtocounter{enumi}{1}
\setcounter{equation}{0}
\renewcommand{\theequation}{\thesection.\theenumi\alph{equation}}
}
\newcommand{\unseteq}{  
\setcounter{equation}{\value{enumi}}
\renewcommand{\theequation}{\thesection.\arabic{equation}}
}

\begin{document}
\hyphenation{anti-fermion}
\topmargin= -20mm
\textheight= 230mm
 \begin{center}
\begin{large}
 {\bf {  STRONG COUPLING LIMIT OF BETHE ANSATZ
         SOLUTIONS IN MASSIVE THIRRING MODEL }}

\end{large}

\vspace{1cm}

T. Fujita\footnote{e-mail: fffujita@phys.cst.nihon-u.ac.jp},
T. Kake,
and H. Takahashi\footnote{e-mail: htaka@phys.cst.nihon-u.ac.jp}

Department of Physics, Faculty of Science and Technology

Nihon University, Tokyo, Japan

\vspace{1cm}

{\Large ABSTRACT}

\end{center}

We study the strong coupling limit of the Bethe ansatz solutions
in the massive Thirring model. We find analytical expressions
for the energy eigenvalues for the vacuum state as well as
$n-$particle $n-$ hole states. This formula is compared
with the numerical results and is found to achieve a very good agreement. 

Also, it is found that the $2-$particle $2-$ hole  and higher
particle$-$hole states
 describe $n-$ free bosons  states in this limit.
The behaviors of the strong coupling limit of the boson mass
for various model calculations are examined. We discuss an ambiguity
of the coupling constant normalization due to the current regularization.

\newpage

\section{Introduction}
Recent calculations for the massive Thirring model have presented
a debate over the energy spectrum of the bound states [1-4]. Several
different  methods give different results on the spectrum
of the bound state. For a long time, people have believed
that the semiclassical results by Dashen et al.[5]
are exact in spite of the fact that they took into account only the
lowest order quantum fluctuations in the path integral method.
However, the recent calculation by the light cone procedure
shows that there is only one bound state, and the spectrum
of the bound state energy as the function of the coupling
constant is different from the semiclassical result [1-3].
Further, the recent calculations  based on the Bethe ansatz solutions [4]
present a numerical proof  that there is only one bound state, and
the spectrum seems to be consistent with the light cone results.

In this paper, we present analytical calculations of the strong coupling
limit of the Bethe ansatz solutions
for the massive Thirring model and show that the analytical expressions
obtained here agree very well with those calculated by  numerically
solving the Periodic Boundary Condition (PBC) equations of the Bethe
ansatz solutions [6,7].
Here, we obtain the energy eigenvalues of the vacuum,
1p$-$1h states (symmetric and asymmetric cases ) and 2p$-$2h and higher particle hole states
(symmetric case). The analytical formula shows that the $n$ particle$-
n$ hole state is just $n$ times 1p$-$1h state energy (a boson mass).
Therefore, there is only one boson state in the massive Thirring model.
This shows that the $n$ particle$-n$ hole states are all scattering 
states.

Further, we show the behaviors of the strong
coupling limit of the boson mass for various model
calculations. It turns out that the analytical expression
of the boson mass at the strong coupling limit with the Bethe ansatz
solutions is different from the
light cone prediction. This may indicate that the normalization ambiguity
of the coupling constant due to the fermion current regularization
in the massive Thirring model is different from the massless Thirring model.
For the massless Thirring model, Klaiber [8] proves that the
coupling constant has a normalization ambiguity which arises from the
fermion current regularization. In the case of the massive Thirring model, 
it is expected that the same type of the coupling constant ambiguity
may well appear. However, we will see later that the normalization 
ambiguity of the coupling constant is more complicated than expected
for the massive Thirring model.

In the next section, we briefly describe the Bethe ansatz method
which is applied to solving the massive Thirring model. 
In section III, we discuss analytical expressions for the energy eigenvalues 
with the strong coupling expansion. 
Then, section IV treats numerical calculations of the PBC equations 
in the strong coupling region and 
we compare them with the analytical expressions. 
In section V, we examine the boson mass at the strong coupling limit. 
Finally, section VI summarizes what we have understood in this paper.

\newpage
\section{Massive Thirring model and Bethe ansatz \protect \\ solutions}

The massive Thirring model is a 1+1 dimensional field theory with
current current interactions [4,11], and can be solved by the Bethe ansatz 
method [4,6,7]. In this case, the eigenvalue problem is reduced to solving 
the following PBC equations, 
\begin{equation}
m_0  \sinh \beta_i = {2\pi n_i \over L} - \sum_j 
{2 \over L} \tan^{-1} \left[{g_B\over 2} \tanh {1\over 2} (\beta_i-\beta_j) \right]
\end{equation}
where $\beta_i$ denotes the rapidity of the particle and $n_i$'s are integer. 
$m_0$, $L$ and $g_B$ denote the bare mass, the box length and the coupling 
constant, respectively.

\subsection{Vacuum state}

The PBC equations 
for the vacuum which is filled with negative energy particles
( $\beta_i=i\pi -\alpha_i$ ) can be written as 
\begin{equation}
\sinh \alpha_i = {2\pi i \over{L_0}} -
 {2\over{L_0}} \sum_{j \not= i} \tan^{-1}\left[{1\over 2}g_B
\tanh{1\over 2} (\alpha_i -\alpha_j) \right],  \qquad
\label{two_vac}
\end{equation}
\[ (i= 0, \pm 1,..,\pm N_0 ) , \]
where $L_0$ is defined as $L_0 = m_0 L$ and $N_0 = (N-1)/2$. 
 In this case, the vacuum energy $E_v$ can be
written as
\begin{equation}
E_v =- \sum_{i=-N_0}^{N_0} m_0 \cosh \alpha_i. 
\end{equation}

\subsection{$1p-1h$ state}

For one particle-one hole $(1p-1h)$
 states, we take out one
negative energy particle ($i_0$-th particle)
 and put it into a positive energy state.
In this case, the PBC equations become
\changeeq
\begin{eqnarray}
i\neq i_0 \nonumber \\
 & \sinh\alpha_i & = \frac{2\pi i}{L_0}-\frac{2}{L_0}\tan^{-1}
 \left[ {g_B\over 2} \coth\frac{1}{2}(\alpha_i+\beta_{i_0})\right]
 \nonumber \\
 & &- \frac{2}{L_0}\sum_{j\neq i,i_0}\tan^{-1}
\left[{g_B\over 2}\tanh\frac{1}{2}
 (\alpha_i-\alpha_j)\right] \\
 \nonumber \\
i=i_0 \nonumber \\
 & \sinh\beta_{i_0} & = \frac{2\pi i_0}{L_0}+\frac{2}{L_0}\sum_{j\neq i_0}
 \tan^{-1}\left[{g_B\over 2}\coth\frac{1}{2}(\beta_{i_0}+\alpha_j)\right] 
\end{eqnarray}
where $\beta_{i_0}$ can be a complex variable as long as it can satisfy
eqs.(2.10).
These  PBC equations determine the energy of the one particle-one
hole states which we denote by $E_{1p1h}^{(i_0)}$,
\unseteq
\begin{equation}
 E_{1p1h}^{(i_0)}= m_0 \cosh \beta_{i_0} -
\sum_{\stackrel{\scriptstyle i=-N_0}{i\not= i_0}}^{N_0}
m_0 \cosh \alpha_i.  
\end{equation}

\subsection{$2p-2h$ states}
For two particle-two hole  $(2p-2h)$ 
states, 
 we take out the $i_1-$th and the $i_2-$th particles
and put them into positive energy states.
The PBC equations for the two particle-two hole states become
\changeeq
\begin{eqnarray}
i\neq i_1, i_2 \nonumber \\
 & \sinh\alpha_i & =  \frac{2\pi i}{L_0}-\frac{2}{L_0}\tan^{-1}
 \left[ {1\over 2}g_B\coth\frac{1}{2}(\alpha_i+\beta_{i_1})\right]
 \nonumber \\
& & -\frac{2}{L_0}\tan^{-1}
 \left[ {1\over 2}g_B\coth\frac{1}{2}(\alpha_i+\beta_{i_2})\right]
 \nonumber \\
 & & - \frac{2}{L_0}\sum_{j\neq i,i_1,i_2}\tan^{-1}
\left[{1\over 2}g_B\tanh\frac{1}{2}
 (\alpha_i-\alpha_j)\right] \\
 \nonumber \\
 i=i_1 \ \ \ \nonumber \\
 & \sinh\beta_{i_1} & = \frac{2\pi i_1}{L_0} +\frac{2}{L_0}\tan^{-1}
\left[{1\over 2}g_B\tanh\frac{1}{2}
 (\beta_{i_1}-\beta_{i_2})\right] \nonumber \\
&&+\frac{2}{L_0}\sum_{j\neq i_1,i_2}
 \tan^{-1}\left[{1\over 2}g_B\coth\frac{1}{2}(\beta_{i_1}+\alpha_j)\right] 
\\ \nonumber \\
i=i_2 \ \ \ \nonumber \\
 & \sinh\beta_{i_2} & = \frac{2\pi i_2}{L_0} + \frac{2}{L_0}\tan^{-1}
\left[{1\over 2}g_B\tanh\frac{1}{2}
 (\beta_{i_2}-\beta_{i_1})\right] \nonumber \\
&&+\frac{2}{L_0}\sum_{j\neq i_1,i_2}
 \tan^{-1}\left[{1\over 2}g_B\coth\frac{1}{2}(\beta_{i_2}+\alpha_j)\right] . 
\end{eqnarray}
In this case, the energy of the $2p-2h$ states  $ E_{2p2h}^{(i_1,i_2)}$
becomes
\unseteq
\begin{equation}
E_{2p2h}^{(i_1,i_2)}= m_0 \cosh \beta_{i_1}+m_0 \cosh \beta_{i_2}
 -\sum_{\stackrel{\scriptstyle i=-N_0}{i\not= i_1,i_2}}^{N_0}
m_0 \cosh \alpha_i . 
\end{equation}
Here, we note that the symmetric case ( $i_1 = -i_2$ ) always gains the 
energy and therefore is lower than other asymmetric cases of $2p-2h$ states.
Higher particle-hole states are constructed just in the same way as above.

\newpage
\section{Strong coupling expansion}
\setcounter{equation}{0}

Here, we present the strong coupling expansion of the PBC equations. 
We take the limit of $g_B \rightarrow \infty$ and obtain the energy 
eigenvalues analytically. This limit of $g_B \rightarrow \infty$ can be 
taken since there is no coupling constant renormalization in the massive 
Thirring model [15].

\subsection{Vacuum state}
First, we consider the strong coupling limit for the vacuum [16]. We assume that
$g_B$ is much larger than any of the rapidity $\alpha_i$, namely,
\begin{equation}
\sqrt{g_B} \gg \alpha_i . 
\end{equation}
In this case, eq.(\ref{two_vac}) becomes in terms of $b_i = \sqrt{g_B L_0} \alpha_i$,
\begin{equation}
b_i=8\sum_{j\neq i}\frac{1}{b_i-b_j}. 
\label{three_beq}
\end{equation}
Further, the vacuum energy is written as
\begin{equation}
E_v=-m_0\sum_{i=-N_0}^{N_0}\cosh\frac{b_i}{\sqrt{g_BL_0}}
 \approx -m_0-2m_0\sum_{i=1}^{N_0}(1+\frac{b_i^2}{2g_BL_0}) . 
\end{equation}
Since $b_i$'s have the symmetry of
$b_i=-b_{-i}$, we can rewrite eq.(\ref{three_beq}) as
\begin{equation}
b_i^2=16\sum_{j=1}^{N_0}\frac{b_i^2}{b_i^2-b_j^2}+12
\ \ \ \ (i=1,\cdots,N_0)   .
\end{equation}
From this equation, we can easily obtain
\begin{equation}
\sum_{i=1}^{N_0}b_i^2=8N_0(N_0-1)+12N_0. 
\end{equation}
Therefore, the vacuum energy can be explicitly written up to
$1/g_B$ order,
\begin{equation}
E_v=-(2N_0+1)m_0-\frac{1}{g_BL}\left[8N_0(N_0-1)+12N_0\right]. 
\label{three_svac}
\end{equation}

\newpage
\subsection{$1p-1h$ state ( symmetric )}
Next, we treat the $1p-1h$ states [15].
We assign the positive energy particle by
$i_0$. In the strong coupling limit, eqs.(2.10) become
\changeeq
\begin{eqnarray}
\sinh\alpha_i &=&  \frac{2\pi i}{L_0}-\frac{\pi}{L_0}
\epsilon(\alpha_i+\beta_{i_0}) \nonumber \\
& & -\frac{2}{L_0}\sum_{j\neq 
i,i_0}\left[\frac{\pi}{2}\epsilon(\alpha_i-\alpha_j)-
\frac{4}{g_B(\alpha_i-\alpha_j)}+...\right] \\
\sinh\beta_{i_0} &=& \frac{2\pi i_0}{L_0} \nonumber \\
& & +\frac{2}{L_0}\sum_{j\neq i_0}
\left[\frac{\pi}{2}\epsilon(\beta_{i_0}+\alpha_j)-
\frac{4}{g_B(\beta_{i_0}+\alpha_j)}+...\right] .
\end{eqnarray}
where $\epsilon(\alpha) $ denote the step function. 
These equations have two solutions, the symmetric and the asymmetric
solutions. For the symmetric case, one easily sees, since
$\alpha_i= -\alpha_{-i}$
\unseteq
\begin{equation}
\beta_{i_0}=0  . 
\end{equation}
Also, for other $\alpha_i$'s,  the equations can be rewritten using $b_i$'s, 
\begin{equation}
b_i^2 =16\sum_{j=1}^{N_0}\frac{b_i^2}{b_i^2-b_j^2}+ 4
 \ \ \ \ (i=1,\cdots,N_0)  .  
\end{equation}
In this case, we can evaluate the energy of $1p-1h$ symmetric case in the same
way as the vacuum and obtain
\begin{equation}
E_{1p1h}^{(0)}=-(2N_0-1)m_0-\frac{1}{g_BL}\left[8N_0(N_0-1)+4N_0\right].
\label{three_s1p1hs}
\end{equation}
Therefore, the $1p-1h$ energy for the $i_0=0$ case
with respect to the vacuum becomes
\begin{equation}
\Delta E_{1p1h}^{(0)}=2m_0+\frac{8N_0}{g_BL}. 
\end{equation}

\subsection{$1p-1h$ states ( asymmetric )}
Now, we discuss the asymmetric solutions. In this case, we obtain
the following coupled equations up to $1/g_B$.
\changeeq
\begin{eqnarray}
\sinh\beta_{i_0} &=& \frac{2\pi }{L_0}(N_0+i_0) -\frac{4}{g_BL_0}
\sum_{j\neq i_0} { \tanh \frac{1}{2}(\beta_{i_0}+\alpha_j) } \\
\sinh\alpha_i &=& \frac{4}{g_BL_0}
 \left[\tanh\frac{1}{2}(\beta_{i_0}+\alpha_i) +\sum_{j\neq {i_0},i}
\frac{1}{\tanh\frac{1}{2}(\alpha_i-\alpha_j)}\right]
\nonumber \\
& &  \quad {\rm for} \quad (N_0 \geq i > i_0) \\
\ \nonumber \\
\sinh\alpha_i &=& -\frac{2\pi}{L_0}+\frac{4}{g_BL_0}
 \left[\tanh\frac{1}{2}(\beta_{i_0}+\alpha_i) +\sum_{j\neq {i_0},i}
\frac{1}{\tanh\frac{1}{2}(\alpha_i-\alpha_j)}\right]
\nonumber \\
& & \quad {\rm for} \quad (i_0-1 \geq i \geq -N_0) .
\end{eqnarray}
From the numerical analysis, we can put
\unseteq
\[
 |\beta_{i_0}| \gg |\alpha_i| . 
\]
Therefore, the above PBC equations are reduced to
\changeeq
\begin{eqnarray}
\sinh\beta_{i_0} &=& \frac{2\pi }{L_0}(N_0+i_0) -\frac{8 N_0}{g_BL_0}
 \tanh \frac{\beta_{i_0}}{2} \label{three_1p1p1}\\
\alpha_i &=& \frac{4}{g_BL_0}
 \left[\tanh\frac{\beta_{i_0}}{2} +\sum_{j\neq {i_0},i}
 \frac{2}{\alpha_i-\alpha_j}\right]
\nonumber \\
& &  \quad {\rm for} \quad (N_0 \geq i > i_0) \label{three_1p1p2}\\
\ \nonumber \\
\alpha_i &=& -\frac{2\pi}{L_0}+\frac{4}{g_BL_0}
 \left[\tanh\frac{\beta_{i_0}}{2} +\sum_{j\neq {i_0},i}
\frac{2}{\alpha_i-\alpha_j}\right]
\nonumber \\
& & \quad {\rm for} \quad (i_0-1 \geq i \geq -N_0) . \label{three_1p1p3}
\end{eqnarray}
The energy of the one particle-one hole states $E_{1p1h}^{(i_0)}$ becomes
\unseteq
\begin{equation}
 E_{1p1h}^{(i_0)} \simeq m_0 \cosh \beta_{i_0} - 2 m_0 N_0
- \frac{m_0}{2} \sum_{\stackrel{\scriptstyle i=-N_0}{i\not= i_0}}^{N_0}
 \alpha_i^2 .
\end{equation}
Therefore, we should calculate the $\cosh\beta_{i_0}$
and the sum of the $\alpha_i^2$. First we evaluate
the $\beta_{i_0}$ from eq.(\ref{three_1p1p1}), which
can be reduced to a cubic equation with
 $x \equiv \tanh(\beta_{i_0}/2)$,
\begin{equation}
 x^3 - \frac{\pi g_B}{4 N_0} (N_0 + i_0) x^2
- \frac{g_B L_0 + 4 N_0}{4 N_0} x + \frac{\pi g_B}{4 N_0} (N_0 + i_0) = 0.
\end{equation}
It is easy to solve the above equation, and 
we obtain 
\begin{equation}
\cosh \beta_{i_0} \simeq \frac{1}{2}
 - \frac{2}{g_B m_0}\left(4 - \pi g_B \frac{N_0+i_0}{N_0}\right) \rho . 
\end{equation}
Next, we calculate the sum of the $\alpha_i^2$. From  eqs.(\ref{three_1p1p2})
and (\ref{three_1p1p3}), we obtain
\begin{eqnarray}
\sum_{\stackrel{\scriptstyle i=-N_0}{i\not= i_0}}^{N_0} \alpha_i^2
 &=& \sum_{i<i_0} \alpha_i^2 + \sum_{i>i_0} \alpha_i^2 \nonumber \\
&=& \left(\frac{2\pi}{L_0}\right)^2 (N_0+i_0)
-\frac{16 \pi}{g_B L_0^2}(N_0+i_0)\tanh\frac{\beta_{i_0}}{2}
-\frac{16 \pi}{g_B L_0^2}\sum_{i<i_0}\sum_{j > i_0}
\frac{2}{\alpha_i-\alpha_j} .
\nonumber \\
\end{eqnarray}
Also, we have from eqs.(\ref{three_1p1p2}) and (\ref{three_1p1p3})
 when $i>i_0$ and $j<i_0$,
\begin{equation}
 \alpha_i - \alpha_j = \frac{2 \pi}{L_0} + O\left(\frac{1}{g_B L_0}\right).
\end{equation}
Therefore, we find
\begin{equation}
\sum_{\stackrel{\scriptstyle i=-N_0}{i\not= i_0}}^{N_0} \alpha_i^2
= \left(\frac{2\pi}{L_0}\right)^2 (N_0+i_0)
-\frac{16 \pi}{g_B L_0^2}(N_0+i_0)\tanh\frac{\beta_{i_0}}{2}
+\frac{16 \pi}{g_B L_0^2}(N_0^2 - i_0^2) .
\end{equation}
Finally, we obtain for the $E_{1p1h}^{(i_0)}$,
\begin{eqnarray}
E_{1p1h}^{(i_0)} &=& m_0 \cosh \beta_{i_0}
- 2 m_0 N_0 - \frac{2 \pi^2}{m_0 L} \left( \rho + \frac{i_0}{L} \right)
\nonumber \\
& & + \frac{8 \pi}{g_B m_0 L} \left( \rho + \frac{i_0}{L} \right)
\tanh \frac{\beta_{i_0}}{2} - \frac{8}{g_B L}(N_0^2 - i_0^2)  .
\label{three_s1p1ha}
\end{eqnarray}

\subsection{$2p-2h$ and higher particle$-$hole states}

Now, we consider $2p-2h$ and higher particle$-$hole states.
In this case, the symmetric solutions always gain the energy.
Therefore, we only treat the symmetric solution here.
Due to the symmetry, we can easily find
\begin{equation}
\alpha_0=0. 
\end{equation}
Let us first consider the $2p-2h$ configuration.
We assume that the $i_0-$th and the $-i_0-$th particles are
in the positive energy state.
In this case, eqs.(2.12) reduce to
\changeeq
\begin{equation}
{b_i^2=16\sum_{j=1}^{N_0-1}\frac{b_i^2}{b_i^2-b_j^2}+12} \ \ \ \
(i\neq i_0),
\end{equation}
\begin{equation}
b_i =-\frac{4}{b_i} \ \ \ \ (i= \pm i_0) . 
\end{equation}
Note that this leads to the $string-$like configurations since the solution for
$b_{i_0}$ becomes pure imaginary. That is,
\unseteq
\begin{equation}
b_{ \pm i_0} = \pm 2i .
\end{equation}
In terms of $\beta_{i_0}$, this becomes
\begin{equation}
\beta_{ \pm i_0} = \pm {2\over{\sqrt{g_B L_0}}}i  . 
\end{equation}
Therefore, the rapidity interval $\Delta$ of the $string$  becomes
\begin{equation}
\Delta =  {4\over{\sqrt{g_B L_0}}}i +
O\left( {i\over{ g_B^{3\over 2}}} \right) ,
\label{three_st}
\end{equation}
where we explicitly write the behavior of the next order of $1/g_B$ expansion.

On the other hand, Bergknoff and Thacker [6] assume that the rapidity
interval of the $string$ behaves for large $g_B$ as
\begin{equation}
 \Delta_{BT} =  {4\over{g_B }}i  .  
\end{equation}
This behavior as the function of $g_B$ does not agree with eq.(\ref{three_st})
which is a solution of the PBC equation. Therefore, the $string$
configurations taken by Bergknoff and Thacker are not consistent with
the $string$-like solution that satisfies the PBC equations.
But this is not at all surprising if one considers the way of obtaining
the $string$ configurations by Bergknoff and Thacker. They assume
that the wave functions of the particles should not diverge at
$x_i = -\infty$ as a sufficient condition. However, this cannot happen
due to the two reasons. The first reason is that one constructs
the field theory in the box of $0 \leq x_i \leq L$. Therefore,
the boundary is always periodic, that is, the wave functions at
$x_i =L$ and $x_i =0$ are the same. The second reason is more physical.
The interactions between particles considered here are always repulsive.
Therefore, the wave functions cannot diverge at any points of the space,
since they are in the scattering states as bare particles.

To avoid the confusions, 
we clarify the $string$ picture. In the nonlinear Schr\"dinger model, 
the $string$ corresponds to the bound states of the  particles since 
they make bound states due to the attractive $\delta-$ type interaction. 
However, this is only possible for the bosonic particles. For fermions, 
there is neither two  particle bound state nor three 
or higher particle bound state
due to the Pauli principle with the $\delta-$ type interaction.
In the massive Thirring model, therefore, we should not consider the $string$
configuration which simulates the many particle bound states.

Now, in the same way as the vacuum case, we obtain
the energy for the $2p-2h$ state
\begin{equation}
E_{2p-2h}^{(0)} =-(2N_0-3)m_0 - \frac{1}{g_BL}
\left[8N_0(N_0-1)-4N_0\right] ,
\end{equation}
where we have ignored those terms which vanish when $L\rightarrow \infty$
with $N_0/L$ finite.
Therefore, the $2p-2h$ energy with respect to the vacuum becomes
\begin{equation}
\Delta E_{2p-2h}^{(0)} = 4m_0+\frac{16N_0}{g_BL}  .
\end{equation}
For higher $p-h$ states, we can evaluate the energy just in
the same way as the $2p-2h$ state case.
For $n$-particle-$n$-hole states, the energy with respect to the vacuum
can be written as
\begin{equation}
\Delta E_{np-nh}^{(0)} =2nm_0+\frac{8nN_0}{g_BL}. 
\end{equation}
It is important to find that the $np-nh$ state energy is just $n$
times as large as that of $1p-1h$ state energy, that is,
\begin{equation}
\Delta E_{np-nh}^{(0)} = n \Delta E_{1p-1h}^{(0)}. 
\end{equation}
This shows that the $n$-particle-$n$-hole states are composed of $n$ free
bosons in this limit.
This result is consistent with the numerical calculations presented in ref.[4].

\section{Numerical method}
\setcounter{equation}{0}

We  solve the PBC equations by the Newton method.
The type of equation we want to solve can be schematically written as
$$  G( {\mbox{\boldmath $f$}} ) =0  \eqno{(4.1)} $$
where ${\mbox{\boldmath $f$}}=(f_1, f_2,..,f_N)$ are the $N$ variables
that should be determined. $G$ is some function.
First, we denote some initial solution by ${\mbox{\boldmath $f_0$}}$.
We expand eq.(4.1) near ${\mbox{\boldmath $f_0$}}$ as
$${\mbox{\boldmath $f$}} = {\mbox{\boldmath $f_0$}}+\delta
{\mbox{\boldmath $x$}} \eqno{(4.2)} $$
$$  G( {\mbox{\boldmath $f_0$}} )+
 {\partial G( {\mbox{\boldmath $f_0$}} )
 \over{\partial {\mbox{\boldmath $f_0$}} }}\delta
{\mbox{\boldmath $x$}} =0  . \eqno{(4.3)} $$
We solve this equation for $\delta {\mbox{\boldmath $x$}} $ and put them
into eq.(4.1). This leads to a new set of ${\mbox{\boldmath $f$}}$, and
we consider ${\mbox{\boldmath $f$}}$ as a new ${\mbox{\boldmath $f_0$}}$
and repeat the same procedure until we get some convergent results
for ${\mbox{\boldmath $f$}}$.

This method has a great advantage over the iteration method
proposed in ref.[4], namely it gives a good convergence {\it even
for the strong coupling region.} However, 
the matrix diagonalization can be possible only for a  few thousand of
matrix dimensions if we have to know all of the eigenvalues. 
It is found that the present calculations 
 agree perfectly with those calculated in ref.[4]. 

It should also be  interesting to check the accuracy of
eqs.(\ref{three_svac}), (\ref{three_s1p1hs}) and (\ref{three_s1p1ha}). 
In Table 1, we show the comparison of the vacuum energies, the $1p-1h$ (symmetric)
and $1p-1h$ (asymmetric) energies between the analytical expressions 
and the numerical calculations with the number of particles of $N=1601$.
As can be seen from the table 1, 
we find quite a good agreement between the predictions
of the strong coupling expansion and the exact numerical calculations.
This indicates that the strong coupling expansion is 
indeed a good approximate scheme.

\vspace{5mm}
\begin{center}
\underline{Table 1} \\
\ \ \\
\begin{tabular}{|c|c|c|}
\hline
\rule{0in}{3.5ex}   & Numerical  &  Analytical \\ \hline
\rule{0in}{3ex}$E_{v}$ & $-1807.44$ & $-1809.08$ \\ \hline
\rule{0in}{3ex}$E_{1p1h}^{(0)}$ & $-1805.20$ & $-1806.81$ \\ \hline
\rule{0in}{3ex}$E_{1p1h}^{(1)}$ & $-1767.40$ & $-1760.44$ \\ \hline
\end{tabular}

\vspace{3mm}
\begin{minipage}{12cm}
The predictions of $1/{g_B}$ expansion are compared with the
computer calculations. Table 1 shows the vacuum energy $E_v$, 
$1p-1h$ energies $E_{1p-1h}^{(0)}$ with the symmetric state 
and $E_{1p-1h}^{(1)}$ with the asymmetric state
for  $g_B = 245$ with $N=1601$ and $L_0 = 100$.
\end{minipage}
\end{center}

\vspace{3cm}
\section{Bound states at the strong coupling limit}
\setcounter{equation}{0}
\subsection{Bound state}
Now, we calculate the bound state of the massive Thirring model at the
strong coupling limit.
In ref.[4], Fujita et al. showed that there is one isolated boson state
and all the other states are continuum states.
Therefore, the 1p-1h continuum energy should start
from the free fermion antifermion mass,
that is twice the physical fermion mass.
\begin{equation}
 m = \frac{1}{2} \left( E^{(1)}_{1p1h}-E_v \right) .
\end{equation}
Thus, the bound state mass ${\cal M}$ can be defined as
\begin{equation}
 {\cal M} = 2m \lim_{\rho \rightarrow \infty}
\left( { \Delta E_{1p1h}^{(0)}\over{\Delta E_{1p1h}^{(1)} }} \right) . 
\label{five_bmass}
\end{equation}
The $\Delta E_{1p1h}^{(1)}$ is given for the large $\rho$
\begin{eqnarray}
\Delta E_{1p1h}^{(1)} &=& E_{1p1h}^{(1)} - E_v \nonumber\\
&=& \frac{3}{2} m_0 +\left[2 \pi - \frac{4}{g_B}
 + \left( \frac{8 \pi}{g_B m_0} - \frac{2 \pi^2}{m_0} \right) \frac{1}{L}
  \right] \rho
\nonumber \\
\end{eqnarray}
Thus, at the large $L$ and $\rho$ limit, $\Delta E_{1p1h}^{(1)}$ becomes
\begin{equation}
\Delta E_{1p1h}^{(1)} = \frac{3}{2} m_0 +\left(2 \pi - \frac{4}{g_B} \right) 
\rho .
\label{five_sd}
\end{equation}
Finally, from eq.(\ref{five_bmass}) and eq.(\ref{five_sd}), we obtain the bound state mass as
\begin{eqnarray}
{\cal M} &=&  2m \lim_{\rho \rightarrow \infty}
\frac{2 m_0 + \displaystyle\frac{8}{g_B} \rho}{\displaystyle\frac{3}{2} m_0 
+   \left(2 \pi - \frac{4}{g_B} \right)\rho } \nonumber \\
 &\simeq& \frac{8}{\pi} \frac{m}{g_B} .
\end{eqnarray}

This result can be compared with the prediction of Fujita and Ogura
in the infinite momentum frame calculation.
\begin{equation}
{\cal M}_{\rm FO}  \simeq 2 \sqrt{2} \pi \frac{m}{g_0}, 
\end{equation}
where $g_0$ is the Schwinger type coupling constant [8,9].
As can be seen, they are different from each other if we
assume $g_B=g_0$.  In ref.[4], it was
shown that the Bethe ansatz solutions for several cases of the coupling
constant are consistent with those of infinite momentum frame
calculation by Fujita and Ogura [1] with the identification of $g_B=g_0$.
However, it became also apparent that the boson mass calculated 
by the Bethe ansatz solution starts to deviate from the light cone result 
in the strong coupling region.
This difference may well be related to the normalization
ambiguity of the coupling constant in the massive Thirring model.

\subsection{Coupling constant ambiguity}

As Klaiber pointed out long time ago, there is an ambiguity
of the coupling constant in the massless Thirring model. 
He proves that the coupling constant is related to each other depending 
on the regularization. We briefly review it in the appendix A.

The results of the boson mass at the strong coupling limit
for the light cone and the Bethe ansatz method indicate that the coupling
constant ambiguity may well be different from eq.(A.8) for the massive
Thirring model. From the comparison of the numerical calculations
and analytical evaluations between the light cone and the Bethe ansatz
solutions, we can anticipate the following relation between $g_0$ and
the Bethe ansatz coupling constant  $g_B$
\begin{equation}
g_B = g_0 \left( {2\sqrt{2}\over{\pi^2}}+
{B\left(1-{2\sqrt{2}\over{\pi^2}}\right)\over{B+{g_0\over{\pi}}}}\right) ,
\label{five_coup}
\end{equation}
where $B$ is a free parameter. In this case, the boson mass of the
Bethe ansatz solutions at the strong
coupling becomes identical to the light cone result.
If we take $B \sim 10$, then
the agreement between the light cone result and the Bethe ansatz solutions 
become very good for whole range of the coupling constant. 
From eq.(\ref{five_coup}), $g_B$ becomes for the small value of $g_0$
\begin{equation}
 g_B \simeq g_0 .
\end{equation}
For the large value of $g_0$, we obtain
\changeeq
\begin{equation}
 g_B \simeq g_0  {2\sqrt{2}\over{\pi^2}} . 
\end{equation}
This can be derived if we assume the following relation for $s$ and $t$
\begin{equation}
t={1\over 2} ({1\over{X}}-1) (s-t) ,
\end{equation}
where $X$ is given as the solution of the following equation,
\unseteq
\begin{equation}
\left[ {g_B\over{\pi}} (B-1)+B- {2\sqrt{2}\over{\pi^2}} \right] X^2
+ \left( {g_B\over{\pi}} -B+{4\sqrt{2}\over{\pi^2}}\right) X-
{2\sqrt{2}\over{\pi^2}} =0  . 
\label{five_coupst}
\end{equation}
Up to now, we do not know any physically simple meaning of choosing the
fermion current regularization which corresponds to eq.(\ref{five_coupst}).
Further studies of the normalization ambiguity of the coupling constant
arising from the fermion current regularization would be very interesting
since we believe that it may well be related to some symmetry
which is hidden in the massive Thirring model.

\newpage
\section{Conclusions}

We have presented numerical calculations as well as the analytical
expressions of the energy eigenvalues of the vacuum and $n$ particle
$n$ hole states in the strong coupling regions.
It is found that the analytical expressions agree very well
with the numerical values
of the vacuum and 1p$-$1h state energies
for the large values of the coupling constant.

From the analytical expressions, we find that the $2p-2h$  and
higher particle hole states appear as free boson states in the strong coupling limit.
This is consistent with the recent
proof [13] that the S-matrix factorization assumed by Zamolodchikov
and Zamolodchikov [14] is violated at the quantum level, and
therefore, the spectrum predicted by the S-matrix factorization is only
semiclassical.

We have also obtained the boson mass ${\cal M}$
at the strong coupling limit analytically.
To compare the present result with other calculations,
we  write here the expressions of the boson mass at the strong coupling 
limit for various calculations.

\renewcommand{\theequation}{6.1\alph{equation}}
\setcounter{equation}{0}
\begin{eqnarray}
 {\cal M}_{\rm DHN} &\simeq& \frac{\pi^2}{2} \frac{m}{g_0} , \\
 {\cal M}_{\rm FO}  &\simeq& 2 \sqrt{2} \pi \frac{m}{g_0} , \\
 {\cal M}_{\rm BA} &\simeq& \frac{8}{\pi} \frac{m}{g_B} ,
\end{eqnarray}
where ${\cal M}_{\rm DHN}$ denotes the result of the
semiclassical calculation by Dashen et al. ,
${\cal M}_{\rm FO}$ is the prediction of the light cone calculation
by Fujita and Ogura, and
the present result is denoted  by ${\cal M}_{\rm BA}$.
If we identify $g_B=g_0$, then ${\cal M}_{\rm BA}$ is different
from the light cone calculation.

At the present stage, we believe that the coupling constant normalization
arising from the fermion current regularization in the massive Thirring
model is slightly different from the massless Thirring model. It may well be 
that the regularization ambiguity is related to some hidden symmetry
which is not clearly understood up to now.

\vspace{1cm}

We would like to thank M. Hiramoto and A. Ogura
for helpful discussions and comments.


\newpage
\appendix
\renewcommand{\theequation}{A.\arabic{equation}}
\setcounter{equation}{0}
\section{Coupling constant ambiguity of the massless Thirring model}
Here, we briefly review
the normalization ambiguity of the coupling constant in the massless
Thirring model [8,12,17].

For the right mover fermion field $\psi_R$, we can express it by
the massless boson fields $\phi_R$ and $\phi_L$ as
\begin{equation}
\psi_R \sim e^{is\phi_R-it\phi_L}
\end{equation}
where $s$ and $t$ are free parameters which satisfy the following constraint 
\begin{equation}
s^2-t^2 = 4\pi . 
\end{equation}
The $s$ and $t$ can be expressed in terms of the boson coupling constant
$\beta$ as
\begin{equation}
s={1\over 2} \left( {4\pi\over{\beta}}+\beta \right), \;\;\;\;
t={1\over 2} \left( {4\pi\over{\beta}}-\beta \right) . 
\end{equation}
Now, the fermion current regularization gives another constraint.
For example, Schwinger's regularization ($g_0$) which makes the fermion current
regularization  only in terms of the space coordinate point splitting
implies that
\begin{equation}
 t={g_0\over{2\pi}} (s-t) .  
\end{equation}
In this case, one obtains the following equation,
\begin{equation}
{\beta^2\over{4\pi}}= {1\over{1+{g_0\over{\pi}}}} . 
\end{equation}
On the other hand, Johnson's regularization ($g$) which makes the fermion current 
regularization  in terms of the space-time coordinate point splitting
in a symmetric fashion implies that
\begin{equation}
 t={g\over{2\pi}} s .  
\end{equation}
In this case, one obtains the following equation,
\begin{equation}
{\beta^2\over{4\pi}}= {2-{g\over{\pi}}\over{2+{g\over{\pi}}}} .
\end{equation}
Therefore, one obtains the relation between $g_0$ and $g$ as written
\begin{equation}
g_0={2g\over{2-{g\over \pi}}} . 
\end{equation}

\end{document}